\documentstyle[preprint,prd,aps,epsfig]{revtex}
\begin{document}
\draft
%%%%%%%%%%%%%%%%% End of Preamble %%%%%%%%%%%%%%%%%%%%%%%
%%%% Start of Text %%%%%%%%%%%%%%%%%%%%%%%%%%%%%%%%%%%%%%%%%%%%%%%%%%%%%%%
\preprint{
\vbox{
\halign{&##\hfil\cr
%	& AS-ITP-98-09 \cr
%       & hep-ph/9806505 \cr
%	& Revised in Oct. 1998 \cr
}}
}
\title{The Pure Leptonic Decays of $B_c$ Meson and Their Radiative Corrections}
\author{Chao-Hsi Chang$^{a, b}$; Cai-Dian L\"{u}$^{b, c}$, Guo-Li Wang$^{b}$; 
Hong-Shi Zong$^{b}$}
\address{$^a$  CCAST (World Laboratory), P.O. Box 8730, Beijing 100080, China}
\address{$^b$  Institute of Theoretical Physics, Academia Sinica, P.O. Box 2735, 
Beijing 100080, China}
\address{$^c$  Department of Physics, Hiroshima University, 1-3-1 Kagamiyama,
739-8526 Higashi-Hiroshima, Japan}
\maketitle
\begin{center}
\begin{abstract}
The radiative corrections to the pure 
leptonic decay $B_{c}{\longrightarrow} 
{\ell}{ {\nu}}_{\ell}$ up-to one-loop order 
is presented. How to cancel the
infrared divergences appearing in the loop calculations, and
the radiative decay $B_{c}{\longrightarrow} 
{\ell}{{\nu}}_{\ell}{\gamma}$ is shown precisely. 
It is emphasized that the radiative decay may be
separated properly and may compare with measurements directly
as long as the theoretical `softness' of the photon corresponds to 
the experimental resolutions. Furthermore
with a kind of non-relativistic constituent quark model, 
a kind of typical long distance contributions to the radiative 
decays is estimated, and it is shown that the contributions
are negligible in comparison with the accuracy of one-loop corrections
and the expected
experimental measurements.

\end{abstract}
\end{center}

\section{Introduction}

\indent
   
The pure leptonic decays of the heavy 
meson $B_c$ are very interesting\cite{cch,du,lcd,aliev,aliev1,lih,gch}. 
In principle, the pure-leptonic 
decay $B_{c}{\longrightarrow} {\ell}{{\nu}_{\ell}}$(see, Fig.1) 
can be used to determine the decay constant $f_{B_{c}}$
if the fundamental Cabibbo-Kobayashi-Maskawa matrix element $V_{bc}$ of 
Standard Model (SM) is known. Conversely if we know the value of decay
constant $f_{B_{c}}$ from other method, these process also can 
be used to extract the matrix element $V_{bc}$. The
$B_{c}$ meson has recently been observed at Fermilab\cite{cdf}, 
that opens a `new page' for the experimental study of $B_c$ meson.
Based on the estimates in Refs.\cite{cch1,k}, numerous $B_{c}$ mesons will 
be produced in Tevatron and also in LHC. Considering the schedule for 
the new runs at Tevatron and LHC constructing, we may expect that experimental 
studies on the $B_{c}$ meson with more care and much higher statistics
will be accessible in the foreseeable future. 

The pure leptonic weak decays of the pseudoscalar meson corresponding to
Fig.1 are helicity suppressed by factor of $m_{\ell}^{2}/m_{B_{c}}^{2}$:
\begin{equation}
\Gamma(B_{c}{\longrightarrow}{\ell}{\overline{\nu}_{\ell}})=
{\frac{G_{F}^{2}}{8\pi}}|V_{bc}|^{2}f_{B_{c}}^{2}m_{B_{c}}^{3}{\frac
{{m_{\ell}}^{2}}{{m_{B_{c}}}^{2}}}
\left(1-{\frac{{m_{\ell}}^{2}}{{m_{B_{c}}}^{2}}}\right)^{2},
\end{equation}
whereas to study them with their radiative decays $B_{c}{\longrightarrow}
{l}{\nu}_l\gamma$ simultaneously is attractive\cite{lcd,aliev,aliev1}.
Of them only the process $B_{c}{\longrightarrow} {{\tau}}{\nu}_{\tau}$ 
is special i.e. it does not suffer so much from the helicity suppression thus  
its branching ratio may reach to $1.5\%$\cite{cch} in SM. However the 
produced $\tau$ will decay promptly and one more neutrino  
is generated in the cascade decay at least, thus it makes 
the decay channel difficult to be observed, especially,
in such a strong background of hadronic collisions.

Fortunately, having an extra real photon emitted in the leptonic decays, 
the radiative pure leptonic decays can escape from the suppression, 
furthermore, as pointed out in \cite{lcd}, with the extra photon 
to identify the produced meson $B_c$ experimentally 
in hadronic collisions from the backgrounds has advantages,
namely in Tevatron and LHC to observe the radiative decays certainly 
has advantages to compare with the pure leptonic ones. 
Although the radiative corrections are suppressed by an
extra electromagnetic coupling constant $\alpha$, it will not be
suppressed by the helicity suppression. Therefore, the
radiative decay may be comparable, even larger than
the corresponding pure leptonic decays. Considering the possibility to
measure the decay constant $f_{B_c}$ and the CKM matrix element
$V_{cb}$ in LHC in the foreseeable future, the problem to increase 
the accuracy of the theoretical calculation, at least up to the first 
order radiative corrections, emerges.

The radiative pure leptonic decays, theoretically, have inferred 
divergences and will be canceled with those from loop corrections of 
the pure leptonic decays, thus we are also interested in 
considering the radiative decays and the pure leptonic decays with
one-loop radiative corrections together. Moreover, all of them
can be divided into two components: 1). The so-called
short distance contributions, for instance, those of the radiative
pure leptonic decays correspond to the four diagrams in
Fig.2, i.e. a real photon is attached to any of the charged lines 
of the pure-leptonic decay Feynman diagrams Fig.1.

2). The so-called long distance contributions which are relevant to 
a virtual heavy hadronic state as an intermediate state. To increase 
the accuracy of the theoretical calculation, both components should
consider precisely.

Recently, there are a few papers\cite{lcd,aliev,aliev1}
on the radiative processes with different methods in literature, 
but inconsistent results have been obtained no matter all of them just 
the lowest order calculations. To clarify the inconsistency,
i.e. further to re-examine the process is certainly needed. Furthermore,
since in Ref.\cite{aliev} an approach of QCD sum rule is adopted 
and in Ref.\cite{lcd} a kind of quark model is adopted, QCD sum rule 
is supposed some long distance effects may be taken into account, so 
it is reasonable to doubt that the disagreement between the results 
of Refs.\cite{lcd,aliev}could be due to the long distance contributions. 
We essentially will adopt the same approach as Ref.\cite{lcd}, thus in 
the paper we especially make some estimates on the long distance 
contributions precisely.

Namely in the present paper, baring the problems and the situation 
pointed out above in mind, we will investigate the $B_{c}$ leptonic decays
up-to one loop radiative corrections carefully. 

The paper is organized as follow: in Section II, we present
the calculations of all the short distance contributions:
{\it i)} besides the leptonic decays shown as Fig.1,
the radiative decays into three family leptons with a real photon 
appearing in the final state as shown in Fig.2; {\it ii)} the virtual photon
corrections to the pure leptonic decays 
i.e. a photon in loop as shown in Figs.3.1, 3.2, 3.3, 3.4. 
In Section III, we estimate a typical
contributions from the long distance contributions. 
In Sec.IV, we evaluate the values of the decays. 
As uncertainties the dependence of the results on 
the parameters appearing in the considered model is
discussed. Comparison with the others' results
are made. Finally, preliminary conclusion is obtained.

%\par
\section{The Short Distance Contributions}

%\indent

Let us start with the radiative decay. The short 
distance contributions are corresponding to the four diagrams 
in Fig.2. According to the constituent quark model which is formulated 
by Bethe-Salpeter (B.-S.) equation, the amplitude turns out to be
the four terms $M_i (i=1,2,3,4)$:
$$M_1=Tr \left[\int\frac{d^{4}q}{(2\pi)^{4}}\chi(p, q) i 
\left(\frac{G_{F}m_{w}^{2}}{\sqrt{2}
}\right)^{\frac{1}{2}}{\gamma}_{\mu}(1-\gamma_{5})V_{bc}\right]\times $$
$${\frac{ i \left(-g^{\mu\nu}+
\frac{p^{\mu}p^{\nu}}{m_{w}^{2}}\right)}{p^{2}-m_{w}^{2}}}
  i e [(p'+p)_{\lambda}g_{\nu\rho}+
(k-p')_{\nu}g_{\rho\lambda}+(-p-k)_{\rho}g_{\nu\lambda} ] 
{\epsilon}^{\lambda}\times$$

\begin{equation}
\frac{ i \left(-g^{{\rho}{\sigma}}+
\frac{(p-k)^{\rho}(p-k)^{\sigma}}{m_{w}^{2}}\right)}{(p-k)^{2}-
m_{w}^{2}}\overline { \ell }\frac{ig}{2\sqrt{2}}\gamma_{\sigma}
(1-{\gamma}_5)\nu_\ell,
\end{equation}

$$M_2=Tr \left [ \int\frac{d^{4}q}{(2\pi)^{4}}\chi(p, q) i 
\left (\frac{G_{F}m_{w}^{2}}{\sqrt{2}
}\right )^{\frac{1}{2}}{\gamma}_{\mu}(1-\gamma_{5})V_{bc}
\right ]{\frac{ i \left(-g^{\mu\nu}+
\frac{p^{\mu}p^{\nu}}{m_{w}^{2}}\right)}{p^{2}-m_{w}^{2}}}$$

\begin{equation}
 \times\overline { \ell }(-ie)\not\! { \epsilon } \frac{i}{\not\! 
 {k}_{\ell}-m_{\ell}}\frac{ig}{2\sqrt{2}}\gamma_{\nu}(1-{\gamma}_5)\nu_\ell,
\end{equation}
 
$$M_3=Tr \left [ \int\frac{d^{4}q}{(2\pi)^{4}}\chi(p, q) 
i\left (\frac{G_{F}m_{w}^{2}}{\sqrt{2}
}\right )^{\frac{1}{2}}{\gamma}_{\mu}(1-\gamma_{5})V_{bc}
\frac{i}{\frac{m_{b}}{m_{b}+m_{c}}\not\! {p}
+\not\! {q}-\not\! {k}-m_{b}}\left(-i \frac{e}{3}\not\! 
{\epsilon}\right)\right ] $$

\begin{equation}
\times\frac{ i \left(-g^{{\mu}{\sigma}}+
\frac{(p-k)^{\mu}(p-k)^{\sigma}}{m_{w}^{2}}\right)}{(p-k)^{2}-
m_{w}^{2}}\overline { \ell }\frac{ig}{2\sqrt{2}}
\gamma_{\sigma}(1-{\gamma}_5)\nu_\ell,
\end{equation}

$$M_4=Tr \left [ \int\frac{d^{4}q}{(2\pi)^{4}}\chi(p, q)
\left( i \frac{2e}{3}\not\! {\epsilon}\right)
\frac{i}{-(\frac{m_{c}}{m_{b}+m_{c}}\not\! {p}
+\not\! {q}-\not\! {k})-m_{c}} i \left (\frac{G_{F}m_{w}^{2}}{\sqrt{2}
}\right )^{\frac{1}{2}}{\gamma}_{\mu}(1-\gamma_{5})V_{bc}\right] $$

\begin{equation}
\times\frac{ i \left(-g^{{\mu}{\sigma}}+
\frac{(p-k)^{\mu}(p-k)^{\sigma}}{m_{w}^{2}}\right)}{(p-k)^{2}-
m_{w}^{2}}\overline { \ell } \frac{ig}{2\sqrt{2}}\gamma_{\sigma}
(1-{\gamma}_5)\nu_\ell,
\end{equation}
where $\chi(p, q)$ is Bethe-Salpeter wave function of the meson $B_{c}$; 
$p$ is the momentum of $B_{c}$; ${\epsilon}, k$ are the polarization 
vector and momentum of the emitted photon.   
In the quark model, the momenta of $\bar {b}, c$-quarks
inside the bound state i.e. the $B_{c}$ meson, are:
%\begin{center}
$$\displaystyle p_{b}={\frac{m_{b}} {m_{b}+m_{c}}}p+q; \;\;\; 
p_{c}={\frac{m_{c}} {m_{b}+m_{c}}}p-q ,$$
%\end{center}
where $q$ is the relative momentum of the two quarks
inside the $B_{c}$ meson. As the $B_c$ meson is a nonrelativistic
bound state in nature, so the higher order relativistic corrections
may be computed precisely, but being an approximation for a $S$-wave
state, and focusing the light on the radiative decay corrections only
at this moment now, we ignore $q$ dependence i.e. we may
still have the non-relativistic 
spin structure for the wave function of the meson $B_c$ (a $^1S_0$ state)
correspondingly:
$$\int\frac{d^{4}q}{(2\pi)^{4}}\chi(p, q)=
\frac{\gamma_{5}({/}\!\!\! {p}{+m})}{2\sqrt{m}}\psi(0).$$ 
Here $\psi(0)$ is the wave function at origin, and by definitions
it connects to the decay constant $f_{B_{c}}$:
$$f_{B_{c}}=\frac{\psi(0)}{2\sqrt {m}},$$
where m is the mass of $B_{c}$ meson. Moreover we note that for 
convenience we take unitary gauge for weak bosons to do the calculations
throughout the paper.
With a straightforward computation, the amplitude can be simplified as:   
$$M_{1}={\frac{-4Ai}{(m^{2}-m_{w}^{2})(m^{2}-2p\cdot k-m_{w}^{2})}}\times$$
\begin{equation}
\overline {\ell}\left (-1+\frac{m^{2}}{m_{w}^{2}}\right )
\left[ p\cdot\epsilon(\not\! {k}-\not\! {p})-(2p\cdot{k}-m^{2})
\not\! {\epsilon} \right](1-\gamma_{5})\nu_{\ell},
\end{equation}

\begin{equation}
M_{2}={\frac{4Ai}{\left(m^{2}-m_{w}^{2}\right)(2k_{1}\cdot{k})}}
\overline {\ell}\left(-1+\frac{m^{2}}{m_{w}^{2}}\right)
\not\! {\epsilon}(\not\! {k}_{1}+\not\! {k}+m_{e})\not\! 
{p}(1-\gamma_{5})\nu_{\ell},
\end{equation}

\begin{equation}
\begin{array}{lcr}
M_{3}+M_{4}&=&\frac{-4Ai}{(-p\cdot{k})(m^{2}-2p\cdot k-m_{w}^{2})}
\overline {\ell}
\left\{\left[-(p\cdot{\epsilon})\not\! {p}+
(p\cdot{\epsilon}\not\! {k}-p\cdot{k}\not\! 
{\epsilon})s_{2} \right.\right.\\ [2mm]
 &+& \left. \left. is_{1}\varepsilon^{\alpha\mu\beta\nu}p_{\alpha}
\epsilon_{\mu}k_{\beta}\gamma_{\nu}\right]+
\frac{(\not\! {p}-\not\! {k})}{m_{w}^{2}}p\cdot{\epsilon}(m^{2}-p\cdot{k})
\right\}(1-\gamma_{5})\nu_{\ell},
\end{array}
\end{equation}
where $k_{1}$ is the momentum of the charged lepton, and
$$A=\frac{\psi(0)\left(\frac{G_{F}m_{w}^{2}}{\sqrt{2}}\right)^{\frac{1}{2}}
V_{bc}eg}{2\sqrt{m}2\sqrt{2}}=
\frac{\psi(0)\left(\frac{G_{F}m_{w}^{2}}{\sqrt{2}}\right)V_{bc}e}{2\sqrt{m}};$$
$$s1=-\frac{m_{b}+m_{c}}{6m_{b}}+\frac{m_{b}+m_{c}}{3m_{c}}; 
\;\; s2=\frac{m_{b}+m_{c}}{6m_{b}}+\frac{m_{b}+m_{c}}{3m_{c}}.$$

\par

As a matter of fact,
there is infrared infinity when performing phase space integral about
the square of matrix element at the soft photon limit. 
It is known that the infrared infinity can be
cancelled completely by that of the 
loop corrections to the corresponding pure leptonic decay $B_c\to l\nu$.

In Eqs.(7) and (8), the infrared terms can be
read out: 
\begin{equation}
M^{i}=M_{2}^{i}+M_{3}^{i}+M_{4}^{i}
=\frac{4Ai}{m_{w}^{2}}\overline{\ell}\left[\frac{k_{1}\cdot{\epsilon}\not\! 
{p}}{k_{1}\cdot{k}}-
\frac{p\cdot{\epsilon}\not\! {p}}{p\cdot{k}}\right](1-\gamma_{5})\nu_{\ell}.
\end{equation}

As the diagrams (g), (h), (i), (j) in Figs.3.3, 3.4 always
have a further suppression factor $m^{2}/{m_{w}}^{2}$
to compare with the other loop diagrams, we may ignore 
the contributions from these four diagrams savely.
Furthermore we should note that in our 
calculations throughout the paper, the dimensional regularization 
to regularize both infrared and ultraviolet divergences is adopted, 
while the on-mass-shell renormalization for the ultraviolet divergence 
is used.

If Feynman gauge for photon is taken (we always do so in the paper), 
the amplitude corresponding to the diagrams (a), (b) of Fig.3.1 
can be written as:

$$M_{(2)}(a)={\frac{2}{3}}eA\int\frac{d^{4}l}{(2\pi)^{4}}
\left[\frac{-4i\varepsilon^{\alpha\mu\beta\nu}p_{\alpha}l_{\beta}-4(p_{\mu}l_{\nu}
-{p\cdot l}g_{\mu\nu}+p_{\nu}l_{\mu})+\frac{8m_{c}}{m_{b}+m_{c}}p_{\mu}p_{\nu}}
{l^{2}(l^{2}-2p\cdot l-m_{w}^{2})(l^{2}-{\frac{2m_{c}}{m_{b}+m_{c}}}p\cdot {l}
)[l^{2}-2l\cdot {(p-k_{2})}]}\right]$$
\begin{equation}
\times\overline{\ell}[2(p-k_{2})_{\mu}-\gamma_{\mu}{\not\!   {l}}]
(-\gamma_{\nu})(1-\gamma_{5})\nu_{\ell},
\end{equation}

$$M_{(2)}(b)={-\frac{1}{3}}eA\int\frac{d^{4}l}{(2\pi)^{4}}
\left[\frac{-4i\varepsilon^{\alpha\mu\beta\nu}p_{\alpha}l_{\beta}+4(p_{\mu}l_{\nu}
-{p\cdot l}g_{\mu\nu}+p_{\nu}l_{\mu})-\frac{8m_{b}}{m_{b}+m_{c}}p_{\mu}p_{\nu}}
{l^{2}(l^{2}-2p\cdot l-m_{w}^{2})(l^{2}-{\frac{2m_{b}}{m_{b}+m_{c}}}p\cdot {l}
)[l^{2}-2l\cdot {(p-k_{2})}]}\right]$$
\begin{equation}
\times\overline{\ell}[2(p-k_{2})_{\mu}-\gamma_{\mu}{\not\! {l}}]
(-\gamma_{\nu})(1-\gamma_{5})\nu_{\ell},
\end{equation}
where the $l$, $k_{2}$ denote the momenta of the loop and the
neutrino respectively. These two terms 
also have infrared infinity when integrating out the loop momentum $l$.
 
After doing the on-mass-shell subtraction, the terms 
corresponding to vertex and 
self-energy diagrams (c), (d), (e), (f) can be written as:
%\begin{center}\begin{eqnarray}
$$M_{(2)}(c+d+e+f)=\frac{ieA}{4\pi^{2}}\overline{\ell}{\not\! p}
(1-\gamma_{5})\nu_{\ell}\times\left[
 ln(4)-\frac{8}{9}+\frac{2}{9}{\frac{m_{b}-m_{c}}{m_{b}+m_{c}}}
ln\left(\frac{m_{b}}{m_{c}}\right)\right.$$%\nonumber\\
 $$+\left(\frac{2}{9}+\frac{8}{9}{\frac{m_{c}}{m_{b}+m_{c}}}\right)
ln\left(\frac{m_{b}+m_{c}}{m_{b}}\right)
 +\left(\frac{8}{9}+\frac{8}{9}{\frac{m_{c}}{m_{b}+m_{c}}}
\right)ln\left(\frac{m_{b}+m_{c}}{m_{c}}\right)$$%\nonumber\\
\begin{equation} +\left.\frac{2}{\varepsilon_{I}}-2\gamma+
ln\left(\frac{4\pi\mu^{2}}{m^{2}}\right)+
 ln\left(\frac{4\pi\mu^{2}}{m_{e}^{2}}\right)\right].
\end{equation}
%\end{eqnarray} \end{center}

Now let us see the cancellation of the infrared divergencies
precisely. The infrared parts of the decay widths
which are from the interference of
the self-energy and vertex correction diagrams
with the tree diagrams:
$$\delta\Gamma_{s,v}^{infrared}=
\displaystyle\left(\frac{\alpha
V_{bc}^{2}f^2_{B_{c}}G^{2}_{F}Mm^{2}_{\ell}}{16{\pi}^2}\right)\left[
-\frac{34}{9}-\frac{4}{9}{\frac{m_{b}-m_{c}}{m_{b}+m_{c}}}ln\left(\frac{m_{b}}
{m_{c}}\right)\right.$$
$$-\left(\frac{2}{9}+\frac{4}{9}{\frac{m_{c}}{m_{b}+m_{c}}}\right)ln\left(\frac{
m_{b}+m_{c}}{m_{b}}\right)
-\left(-\frac{4}{9}+\frac{4}{9}{\frac{m_{b}}{m_{b}+m_{c}}}\right)ln\left(\frac{m
_{b}+m_{c}}{m_{c}}\right)$$
\begin{equation}
-\left.\frac{2}{\varepsilon_{I}}+2\gamma-ln\left(\frac{4\pi\mu^{2}}{m^{2}}\right
)-
 ln\left(\frac{4\pi\mu^{2}}{m_{e}^{2}}\right)\right] \; ;
\end{equation}
the infrared part of the decay widths from the interference
of the "box" correction diagrams with the tree diagrams:
\begin{equation}
\delta\Gamma_{box}^{infrared}=
\left(\frac{\alpha V_{bc}^{2}f^2_{B_{c}}G^{2}_{F}Mm^{2}_{\ell}}
{16{\pi}^2}\right)\left[
-\left(\frac{1}{\varepsilon_{I}}-\gamma\right)ln\left(\frac{m^2_{\ell}}{M^2}
\right)-
ln\left(\frac{m^2_{\ell}}{M^2}\right)ln\left(\frac{4\pi\mu^2}{M^2}\right)\right]\; ;
\end{equation}
the infrared part of the decay width from the real photon emission:
$$ \delta\Gamma_{real}^{infrared}= 
\left(\frac{\alpha V_{bc}^{2}f^2_{B_{c}}G^{2}_{F}Mm^{2}_{\ell}}
{16{\pi}^2}\right)$$
\begin{equation}
\cdot \left[\frac{2}{\varepsilon_I}-2\gamma+\left(\frac{1}{\varepsilon_{I}}-
\gamma\right)ln\left(\frac{m^2_{\ell}}{M^2}\right)
+\left( 2+ ln\left( \frac{m^2_{\ell}}{M^2} \right)
\right) ln \left(\frac{4\pi\mu^2}{4(\Delta E)^2} \right) \right] \;.
\end{equation}
Here $\Delta{E}$ is a small energy, which corresponds to the
experimental resolution of a soft photon so that the phase space of the
emitting photon in fact is divided into a soft and a hard part
by $\Delta{E}$. Where $\mu$ is the dimensional parameter 
appearing in the dimensional regularization. 
It is easy to check that when adding up all the parts: the real photon
emission $\delta\Gamma_{real}^{infrared}$ and the virtual photon 
corrections $\delta\Gamma_{s,v}^{infrared}, \delta\Gamma_{box}^{infrared}$, 
the infrared divergences $\frac{2}{\varepsilon_{I}}-\gamma$ are canceled 
totally and the $\mu$ dependence is also cancelled. Hence we may be sure 
that we finally obtain the pure leptonic widths for the $B_c$ meson decays 
to the three families of leptons which are accurate up-to the `next'-order 
corrections and `infrared finite' but depend on the experimental resolution
$\Delta{E}$. 
 
\section{Estimate of the Long Distance Contributions}
 
We have done the calculations of the radiative `short distance' corrections
corresponding to the energy scale around $m_{B_c}$, whereas, there are
corrections which correspond to much softer nature than what we have
considered. Typically, some of them in the radiative 
decay $B_c \rightarrow l\nu\gamma$ may be described by Fig.4. 
People, generally, call the soft corrections as the `long 
distance contributions' correspondingly.

To have a rough estimate on the contributions
of the softer part and to be typical, let us only consider 
those corresponding to Fig.4. 

The amplitude for the long distance radiative corrections 
corresponding to Fig.4 may be written as:
$$M={\frac{-ie(2\pi)^{4}\delta^{4}(P_{\small{B}_{c}}-k-k_{1}-k_{\nu})
G_{F}\epsilon_{\mu}(k)}{\sqrt 2}}\overline{\ell}(k_{1})\gamma^{\lambda}
(1-\gamma_{5})\nu_{e}(k_{\nu})$$
\begin{equation}
\cdot \left[\sum\limits_{\stackrel{\rightarrow}{p_{n}}=\stackrel{\rightarrow}{k}}
\frac{\langle{0}|j^{\mu em}(0)|p_{n}\rangle\langle{p_{n}}|j_{q\lambda}^{+}(0)|P_
{B_{c}}\rangle}{2p_{n0}(k_{0}-p_{n0}+i\epsilon)}+
\sum\limits_{\stackrel{\rightarrow}{p_{n}}=\stackrel{\rightarrow}
{P_{\small{B}_{c}}}-\stackrel{\rightarrow}{k}}
\frac{\langle{0}|j_{q\lambda}^{+}(0)|p_{n}\rangle\langle{p_{n}}|j^{\mu em}(0)|P_
{\small{B}_{c}}\rangle}{2p_{n0}(P_{\small{B}_{c0}}-k_{0}-p_{n0}+i\epsilon)}
\right],\\[2mm]
\end{equation}
where $j^{\mu em}$, $j_{q\lambda}^{+}$ 
are electromagnetic current and 
weak current respectively, and $\epsilon(k)$ 
is polarization vector of 
the real photon. The intermediate states $|p_{n}\rangle$ are 
all the possible physical states, 
and in the summation 
$\stackrel{\rightarrow}{p_{n}}=\stackrel{\rightarrow}{k}$ for
the first term and $\stackrel{\rightarrow}{p_{n}}=
\stackrel{\rightarrow}{P_{B_{c}}}-\stackrel{\rightarrow}{k}$ for the
second term are kept. Note that for the time-component, generally,
$k_{0}\neq p_{n0}$ for the first term and
$p_{n0}\neq P_{B_{c0}}-k_{0}$ for the second term.
To compute the whole amplitude, let us compute 
the current matrix elements appearing in Eq.(16) 
with the intermediate meson states being on mass-shell. 
However to have a rough estimate instead of a precise one, here
only the intermediate meson states 
$J/\psi$ and $B_{c}^{*}$, being of typical long distance contributions,
are computed by means of the three Feynman diagrams shown in Fig.4.

According to Eq.(16), we have:
\begin{equation}
\langle{0}|j^{\mu em}(0)|p_{J/\psi}\rangle=-{\frac{2}{3}}{\frac{1}
{\sqrt{2\pi p_{J/\psi 0}}}}\phi_{J/\psi}(0)M_{J/\psi}\epsilon_{\mu}(J/\psi),
\end{equation}

\begin{equation} 
   \langle{0}|j_{q\lambda}^{+}(0)|p_{B^{*}_{c}}\rangle=-{\frac{1}
{\sqrt{2\pi p_{B^{*}_{c} 0}}}}\phi_{B^{*}_{c}}(0)
M_{B^{*}_{c}}\epsilon_{\mu}(B^{*}_{c}).
\end{equation}
The matrix element $\langle p_{J/\psi}|j_{q\lambda}^{+}(0)|P_{B_{c}}\rangle$
can be decomposed into two parts:
\begin{equation}
\langle p_{J/\psi}|j_{q\lambda}^{+}(0)|P_{B_{c}}\rangle=
\langle p_{J/\psi}|V_{\lambda}|P_{B_{c}}\rangle
-\langle p_{J/\psi}|A_{\lambda}|P_{B_{c}}\rangle,
\end{equation}
namely those of the vector current $V_{\lambda}={\frac{1}{2}}\overline{c}
\gamma _{\lambda}b$
and the axial current $A_{\lambda}={\frac{1}{2}}\overline{c}
\gamma _{\lambda}\gamma_{5}b$. They
are related to the form factors\cite{cch,s11,s12} as follows: 

\begin{equation}
\langle p_{J/\psi}|V_{\mu}|P_{B_{c}}\rangle=
ig{\frac{1}{2}}\varepsilon_{\mu\nu\rho\sigma}
\epsilon^{*\nu}(P_{B_{c}}+p_{J/\psi})^{\rho}(P_{B_{c}}-p_{J/\psi})^{\sigma},
\end{equation}
\begin{equation}
\langle p_{J/\psi}|A_{\mu}|P_{B_{c}}\rangle={\frac{1}{2}}[f\epsilon^{*}_{\mu}+a_{+}
(\epsilon^{*}\cdot P_{B_{c}})(P_{B_{c}}+p_{J/\psi})_{\mu}+a_{-}
(\epsilon^{*}\cdot P_{B_{c}})(P_{B_{c}}-p_{J/\psi})_{\mu}],
\end{equation}
where $g, f, a_+, a_{-}$ are the possible form factors. 
For convenience, here we actually calculate
the matrix element $\langle p_{B_{c}}|j^{\mu em}(0)|P_{B_{c}^{*}}\rangle$ 
instead of $\langle p_{B_{c}^{*}}|j^{\mu em}(0)|P_{B_{c}}\rangle$:
\begin{equation}
\langle p_{B_{c}}|j^{\mu em}(0)|P_{B^{*}_{c}}\rangle=i(c_{b}g_{b}+c_{c}g_{c})
\varepsilon_{\mu\nu\rho\sigma}\epsilon^{*\nu}({B_{c}^{*}})(P_{B_{c}^{*}}+
p_{B_{c}})^{\rho}(P_{B_{c}^{*}}-p_{B_{c}})^{\sigma}
\end{equation}

Now we just present the final results of the form factors $g, f, a_{+}, 
a_{-}, g_{b}, g_{c}$ here, as the
detail calculations on them can be found in Ref.[1]. 

Let us introduce further definitions, which is convenient for presenting
the results. If $p_{1}, p_{2}$ are the momenta of the constituent 
particles 1 and 2 respectively,
the total and the relative momenta $p$ 
and $q$ are defined as:
%\begin{center}
$$p_{1}={\alpha}_{1}p+q, \;\; {\alpha}_{1}=\frac{m_{1}}{m_{1}+m_{2}};$$
$$p_{2}={\alpha}_{2}p+q, \;\; {\alpha}_{2}=\frac{m_{2}}{m_{1}+m_{2}}.$$
%\end{center}
The momenta $p$, $p'$ are replaced to denote those for the
initial and final mesons $P_{B_{c}}$, $p_{B_{c}^{*}}$ (or $p_{J/\psi}$), 
and $M, M'$ are denoted the masses of the initial and final mesons.

Furthermore $q_{p}$, $q_{pT}$ are denoted the two Lorentz covariant variables
as follows:
$$q_{p}={\frac{p\cdot q}{M_{p}}}, \;\;\; q_{pT}=\sqrt{q^{2}_{p}-q^{2}}$$
and:
$$\omega_{ip}=\sqrt{m_{i}^{2}+q^{2}_{pT}}$$
$${{\omega}'}_{ip'}=\sqrt {{m'}_{i}^{2}+{q'}^{2}_{p'T}}.$$

Now we can give the formulas of form factor using the above covariant 
variables:
$$ g=\xi{\frac{\omega_{1}^{'}+\omega_{2}^{'}}{MM^{'}m_{1}^{'}}} $$
$$ f=\xi \left[{\frac{(p\cdot p_{1}^{'})}{Mm_{1}^{'}}}+1\right] $$
\begin{equation}
a_{\pm}=\xi\left\{{\frac{2m_{2}}{M^{2}m'_{1}}}+
\delta\mp\left[{\frac{{\omega}'_{1}+{\omega}'_{2}}
{MM'm'_{1}}}+{\frac{(p\cdot{p'})}{M'^{2}}}\delta\right]\right\}
\end{equation}
where
$$\delta=-{\frac{C[1+(p\cdot{p'_{1}})/Mm'_{1}]}{p^{2}-(p\cdot{p'})^{2}/M'^{2}}}$$
$$C={\frac{1}{L_{+}\left[{\frac{({p'}\cdot p_{1})({p'}
\cdot {p'}_{1})}{{M'}^{2}}}-{\frac{1}{L_{-}^{2}}}\right]^{1/2}-1}}$$
where
$$L_{\pm}=\left[{\frac{1}{2}}(p_{1}\cdot {p'}_{1}\pm m_{1}{m'}_{1} )
\right]^{-1/2}.$$
The common factor:
\begin{equation}
\xi=\left[{\frac{2{\omega}'_{2}m^{2}_{1}m'^{2}_{1}}{[(p_{1}
\cdot{p'_{1}})+m_{1}m'_{1}]
{\omega}_{1}{\omega}'_{1}{\omega}_{2}}}\right]^{1/2}
\times\int\frac{d^{3}\stackrel\rightarrow
{q}}{(2\pi)^{3}}{\phi}'^{*}_{p'}(q'_{p'_{T}})
\cdot{\phi_{p}(|\stackrel
{\rightarrow}{q}|)}
\end{equation}
if it is written in c.m.s. of the initial meson
($\stackrel{\rightarrow}{p}=0$).

%\indent
Here ${\phi}'^{*}_{p'}(q'_{p'_{T}})$ and ${\phi_{p}(|\stackrel
{\rightarrow}{q}|)}$\cite{s14} correspond to the radius wave functions
of the mesons in the initial and final states respectively, but both are 
presented in c.m.s. of the initial meson.
The equations about $g_{a}$, $g_{b}$ are similar as $g$, we will not repeat 
them here.

\section{Numerical Results and Discussion}

%\indent 
First of all, let us show the `whole' leptonic decay widths
i.e. the sum of the corresponding radiative decay widths 
and the corresponding pure leptonic decay widths
with radiative corrections, and put them in Table (1). As 
firstly we would like 
to see the facts of `short distance' contributions, 
so here only the `short 
distance' contributions are taken into account. 
Why we put the radiative decay 
and the pure leptonic decay with radiative corrections together here is to
make the width not to depend on the experimental resolution for a soft photon
at all. To compare with the earlier computations\cite{lcd,aliev,aliev1} 
in the numerical evaluation, the values for the parameters $\alpha=1/132$ and 
$|V_{bc}|=0.04$\cite{data} are taken, and two possible values for
the rests are selected as bellow:\\
(1)  $m_{B_{c}}=6.258$ GeV, $m_{b}=4.758$ GeV, $m_{c}=1.500$ GeV,
$f_{B_{c}}=0.360$ GeV\cite{lcd};\\
(2)  $m_{B_{c}}=6.258$ GeV, $m_{b}=4.700$ GeV, $m_{c}=1.400$ GeV,
$f_{B_{c}}=0.350$ GeV\cite{aliev,aliev1}.

\begin{center}
Table (1) The `Whole' Leptonic Decay Widths (in unit GeV)\\ 
\noindent
{\small (short distance contributions only)}\\
\vspace{2mm}
\begin{tabular}{|c|c|c|} \hline
  &(1) &(2)  \\  \hline
 $\Gamma_{e}(10^{-17})$ &   6.444 & 6.902 \\ \hline
 $\Gamma_{\mu}(10^{-16})$ &  1.383 & 1.389 \\ \hline
 $\Gamma_{\tau}(10^{-14})$ & 1.871 & 1.782 \\\hline
\end{tabular}
\end{center}

For comparison, the width of each pure leptonic decay 
at tree level with the same parameters as those in Table (1) is 
put in Table (2).
                               
\begin{center}
Table (2) The Pure Leptonic Decay Widths (in unit GeV) of Tree Level\\
\vspace{2mm}
\begin{tabular}{|c|c|c|} \hline
  &(1)&(2)
    \\  \hline
 $\Gamma_{e}(10^{-21}$)& 1.827& 1.727 \\ \hline
 $\Gamma_{\mu}(10^{-16}$) &  0.7841& 0.7412\\ \hline
 $\Gamma_{\tau}(10^{-14}$)&  1.862& 1.773 \\ \hline
\end{tabular}
\end{center}

If the lifetime of $B_{c}$ meson is (a). $\tau(B_{c})=0.46\times10^{-12}s$
as indicated by the first observation\cite{cdf}; (b). $\tau(B_{c})=0.52\times
10^{-12}s$ as adopted in Ref.\cite{lcd}, the corresponding branching ratios 
are showed in Tables (3), (4).

\begin{center}
Table (3) Branching Ratios of the `Whole' Leptonic Decays \\
\noindent
{\small (short distance contributions)}\\
\vspace{2mm}
\begin{tabular}{|c|c|c|c|c|} \hline
  &(1-a)&(2-a)&(1-b)&(2-b)
    \\  \hline
 $B_{e}(10^{-5})$ & 5.09 & 5.45&4.5&4.82\\ \hline
 $B_{\mu}(10^{-5})$ & 10.93& 10.98 & 9.69 & 9.76 \\ \hline
 $B_{\tau}(10^{-2})$& 1.477 & 1.407 & 1.306 &1.246 \\ \hline
\end{tabular}
\end{center}

\begin{center}
Table (4) Tree Level Branching Ratios of The Pure Leptonic Decays\\
\vspace{2mm}
\begin{tabular}{|c|c|c|c|c|} \hline
  &(1-a)&(2-a)&(1-b)&(2-b)
    \\  \hline
 $B_{e}(10^{-9})$&1.44&1.36&1.28&1.21\\ \hline
 $B_{\mu}(10^{-4})$ &0.62&0.586&0.55&0.52\\ \hline
 $B_{\tau}(10^{-2})$&1.47&1.40&1.30&1.24\\ \hline
\end{tabular}
\end{center}

To see the contributions of the radiative decays precisely
we present the radiative decay widths with a cut of the
photon energy i.e. the widths of the radiative decays 
$B_c\rightarrow l\nu\gamma$ with the photon energy $E_\gamma \geq k_{min}$
as the follows: $k_{min}=0.1$ GeV, $k_{min}=0.2$ GeV, 
$k_{min}=0.5 $ GeV and $k_{min}=1.0 $ GeV respectively in Table (5).

\begin{center}
Table (5): The Radiative Decay Widths (in unit $10^{-17}$GeV)\\ 
\noindent
{\small (with cuts of the photon momentum and the angle 
between photon and lepton)}\\
\vspace{2mm}
\begin{tabular}[t]{|c|c|c|c|}
\hline
&  & (1) & (2) \\ \hline
$k_{min(GeV)}$ &  & $\quad \;5^0\quad \quad \;\;\;15^0\quad \quad
\;\;30^0\,\quad $ & ${}{}5^0\;\;\qquad \;15^0\;\;\qquad 30^0$ \\ \hline
$0.1$ & $\Gamma _e$ & 6.384\qquad 6.370\qquad 6.297 & \ 6.832\qquad
6.819\qquad 6.752\  \\ \hline
$0.2$ & $\Gamma _e$ & \thinspace 6.317\qquad 6.303\qquad 6.242 & 6.762\qquad
6.750\qquad 6.693 \\ \hline
$0.5$ & $\Gamma _e$ & 5.931\qquad 5.918\qquad 5.883 & 6.351\qquad
6.340\qquad 6.307 \\ \hline
$1.0$ & $\Gamma _e$ & 4.807\qquad 4.800\qquad 4.790 & 5.151\qquad
5.143\qquad 5.136 \\ \hline
$0.1$ & $\Gamma _\mu $ & 6.613\qquad 6.518\qquad 6.385 & 7.049\qquad 
6.958\qquad 6.834 \\ \hline
$0.2$ & $\Gamma _\mu $ & 6.484\qquad 6.412\qquad 6.306 & 6.920\qquad
6.850\qquad 6.753 \\ \hline
$0.5$ & $\Gamma _\mu $ & 6.018\qquad 5.977\qquad 5.917 & 6.433\qquad
6.394\qquad 6.340 \\ \hline
$1.0$ & $\Gamma _\mu $ & 4.843\qquad 4.824\qquad 4.802 & 5.184\qquad
5.165\qquad 5.146 \\ \hline
$0.1$ & $\Gamma _\tau $ & 13.75\qquad 13.66\qquad 12.88 & 13.60\qquad
13.52\qquad 12.78 \\ \hline
$0.2$ & $\Gamma _\tau $ & 10.87\qquad 10.82\qquad 10.34 & 10.86\qquad
10.81\qquad 10.36 \\ \hline
$0.5$ & $\Gamma _\tau $ & 7.139\qquad 7.121\qquad 6.970 & 7.282\qquad
7.265\qquad 7.122 \\ \hline
$1.0$ & $\Gamma _\tau $ & 4.169\qquad  4.165\qquad 4.146 & 4.340\qquad
4.335\qquad 4.318 \\ \hline
\end{tabular}
\end{center}

\medskip

For the convenience to compare with experiments, we present the photon
spectrum of the radiative decays in Fig.6 
and give the decay widths
of the so-called physical pure leptonic decays: 
$\Gamma _{phys}=\Gamma _{whole}-\Gamma _{cut}$ 
where $\Gamma _{cut}$ are showed in Table (5); $\Gamma _{whole}$
are showed in Table (1). We put the values of $\Gamma _{phys}$ in Table (6).

\begin{center}
Table (6): Decay widths of $\Gamma _{phys}$ ($10^{-17}$GeV) \\
\vspace{2mm}
\begin{tabular}{|c|c|c|c|}
\hline
&  & (1) & (2) \\ \hline
$k_{min}$(GeV) &  & $5^0\;\qquad \;\;\;15^0\qquad \;\;30^0$ & $5^0\;\qquad
\;\;15^0\;\qquad \;30^0$ \\ \hline
$0.1$ & $\Gamma _e$ & 0.060\qquad 0.074\qquad 0.147 & 0.070\qquad
0.083\qquad 0.150 \\ \hline
$0.2$ & $\Gamma _e$ & 0.127\qquad 0.141\qquad 0.202 & 0.140\qquad
0.152\qquad 0.209 \\ \hline
$0.5$ & $\Gamma _e$ & 0.513\qquad  0.526\qquad 0.561 & 0.551\qquad
0.562\qquad 0.595 \\ \hline
$1.0$ & $\Gamma _e$ & 1.637\qquad 1.644\qquad 1.654 & 1.751\qquad
1.759\qquad 1.766 \\ \hline
$0.1$ & $\Gamma _\mu $ & 7.217\qquad 7.312\qquad 7.445 & 6.837\qquad
6.928\qquad 7.052 \\ \hline
$0.2$ & $\Gamma _\mu $ & 7.346\qquad 7.418\qquad 7.524 & 6.966\qquad
7.036\qquad 7.133 \\ \hline
$0.5$ & $\Gamma _\mu $ & 7.812\qquad 7.853\qquad 7.913 & 7.453\qquad
7.492\qquad 7.546 \\ \hline
$1.0$ & $\Gamma _\mu $ & 8.987\qquad 9.006\qquad 9.028 & 8.702\qquad
8.721\qquad 8.740 \\ \hline
$0.1$ & $\Gamma _\tau (10^3)$ & 1.857\qquad 1.857\qquad 1.858 & 1.768\qquad
1.768\qquad 1.769 \\ \hline
$0.2$ & $\Gamma _\tau (10^3)$ & 1.860\qquad 1.860\qquad 1.860 & 1.771\qquad
1.771\qquad 1.772 \\ \hline
$0.5$ & $\Gamma _\tau (10^3)$ & 1.864\qquad 1.864\qquad 1.864 & 1.775\qquad
1.775\qquad 1.775 \\ \hline
$1.0$ & $\Gamma _\tau (10^3)$ & 1.866\qquad 1.866\qquad 1.867 & 1.777\qquad
1.777\qquad 1.778 \\ \hline
\end{tabular}
\end{center}

Let us select the parameters as in Ref.\cite{lcd}: $m_{B_c}=6.258$
GeV, $m_b=4.758$ GeV, $m_c=1.500$ GeV, $f_{B_c}=0.360$ GeV, 
so as to compare with the short
distance contributions and to show the contributions 
from the typical long distance component
concerned in the paper. For each of the family, the contribution
to the width:
$$\delta\Gamma _e=1.828\times10^{-18} GeV,$$
$$\delta\Gamma _\mu =1.827\times10^{-18} GeV,$$
$$\delta\Gamma _\tau =1.190\times10^{-18} GeV.$$

From Table (5), we my see that the present results confirm those 
in Ref.\cite{lcd}, and the slight differences are maily due to the
radiative corrections and the
fact that instead of ignoring in Ref.\cite{lcd}, we keep the lepton mass
precisely in the numerical evaluation. In comparison 
with the short contributions, for the radiative leptonic decays 
of the $B_{c}$ meson the long distance contributions 
$\delta\Gamma _l, (l=e,\mu,\tau)$
are quite small. It may be the reason that the $b$ and $c$ quark in 
$B_{c}$ meson are heavy. Their heavy mass causes the long distance 
contributions small. Therefore, it seems that we can conclude that the
disagreement between Ref.\cite{lcd} and Refs.\cite{aliev,aliev1} is not
due to the long distance effects.

In addition, we should note that the widthes are quite sensitive to 
the decay constant $f_{B_{c}}$, and the values of the quark masses 
$m_b$ and $m_c$.

Note that when we almost completed this paper, one similar 
paper\cite{s17} appeared and its results supported those
in Ref.\cite{lcd} that is in agreement with
this paper.

\vspace{2cm}
{\Large\bf Acknowledgement} This work was supported in part by the National 
Natural Science Foundation of China and the Grant No. LWLZ-1298 of the Chinese 
Academy of Sciences. One of the author (C.-D. L\"u) would 
also like to thank JSPS for supporting him in research.

%\end{thebibliography}

\begin{figure}\begin{center}
   \epsfig{file=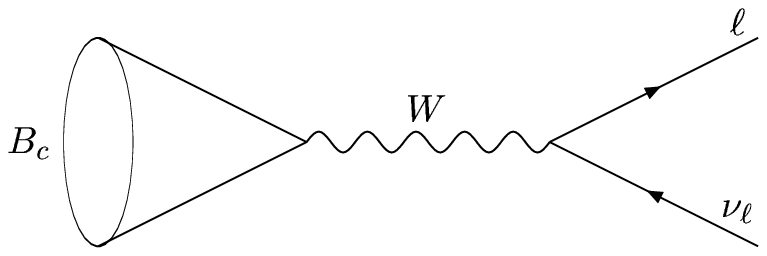, bbllx=146pt,bblly=300pt,bburx=371pt,bbury=393pt,
width=10cm,angle=0}
\caption{\bf Tree diagram for $B_{c}\longrightarrow\ell\nu_{\ell}$.}
%\label{fig}
\end{center}
\end{figure}

\begin{figure}\begin{center}
   \epsfig{file=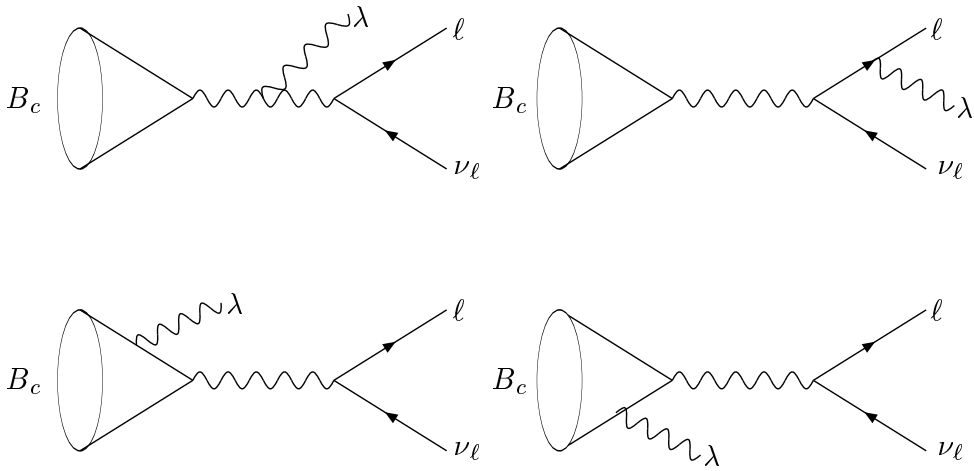, bbllx=157pt,bblly=497pt,bburx=505pt,bbury=683pt,
width=13cm,angle=0}
\caption{\bf Diagrams for $B_c\longrightarrow \ell \nu \gamma $.}
%\label{fig}
\end{center}
\end{figure}
\setcounter{figure}{2}

\begin{figure}\begin{center}
   \epsfig{file=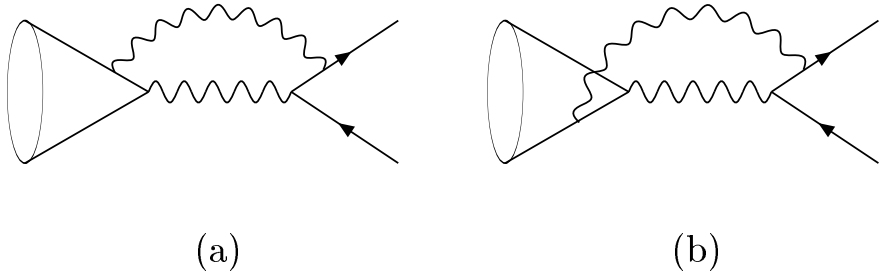, bbllx=146pt,bblly=230pt,bburx=400pt,bbury=316pt,
width=12cm,angle=0}
\caption{\bf 1. Box-loop diagrams for $B_c \longrightarrow \ell \nu$.}
%\label{fig}
\end{center}
\end{figure}

\setcounter{figure}{2}
\begin{figure}\begin{center}
   \epsfig{file=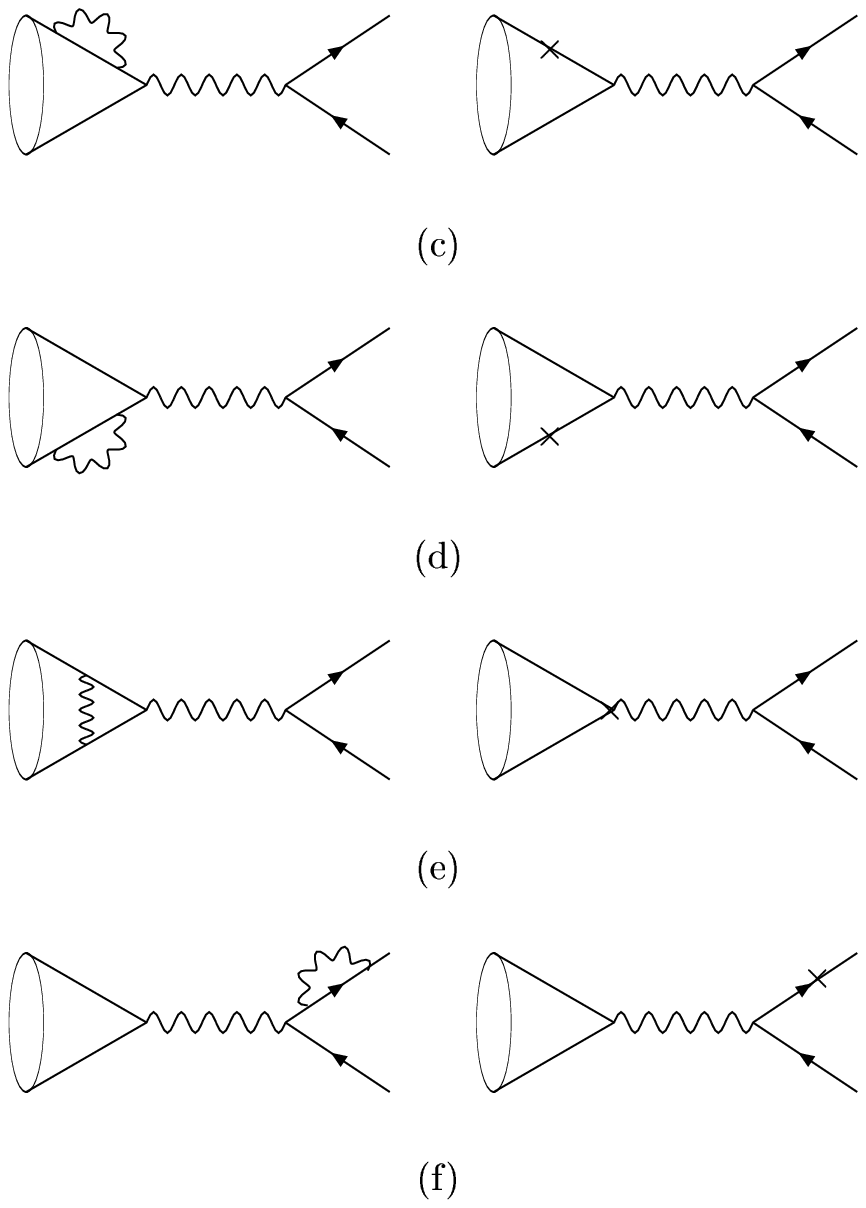, bbllx=277pt,bblly=312pt,bburx=536pt,bbury=676pt,
width=12cm,angle=0}
\caption{\bf 2. Self-energy and vertex diagrams for $B_c \longrightarrow \ell \nu$.}
%\label{fig}
\end{center}
\end{figure}

\setcounter{figure}{2}
\begin{figure}\begin{center}
\epsfig{file=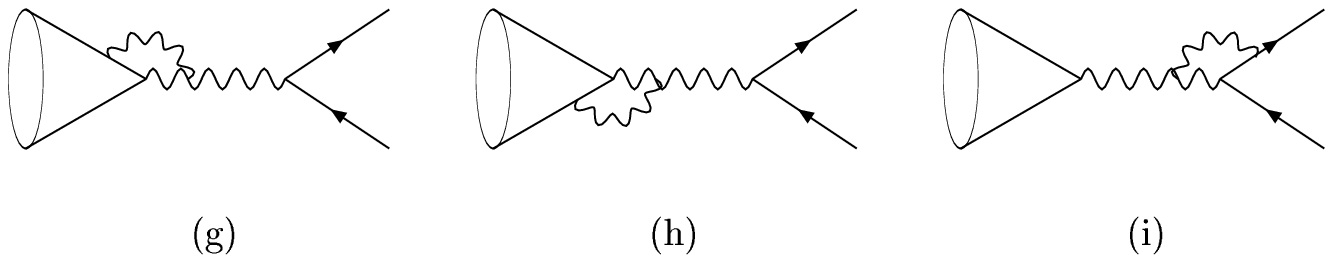, bbllx=146pt,bblly=234pt,bburx=530pt,bbury=312pt,
width=14cm,angle=0}
\caption{\bf 3. Vertex diagrams for $B_c \longrightarrow \ell \nu$.}
%\label{fig}
\end{center}
\end{figure}

\setcounter{figure}{2}
\begin{figure}\begin{center}
\epsfig{file=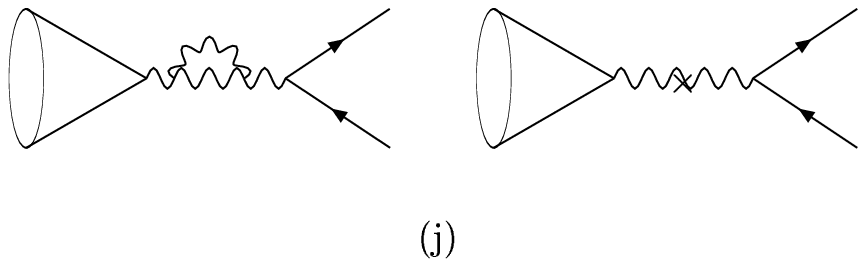, bbllx=277pt,bblly=535pt,bburx=530pt,bbury=610pt,
width=12cm,angle=0}
\caption{\bf 4. Self-energy diagrams for $B_c \longrightarrow \ell \nu$.}
%\label{fig}
\end{center}
\end{figure}

\begin{figure}\begin{center}
   \epsfig{file=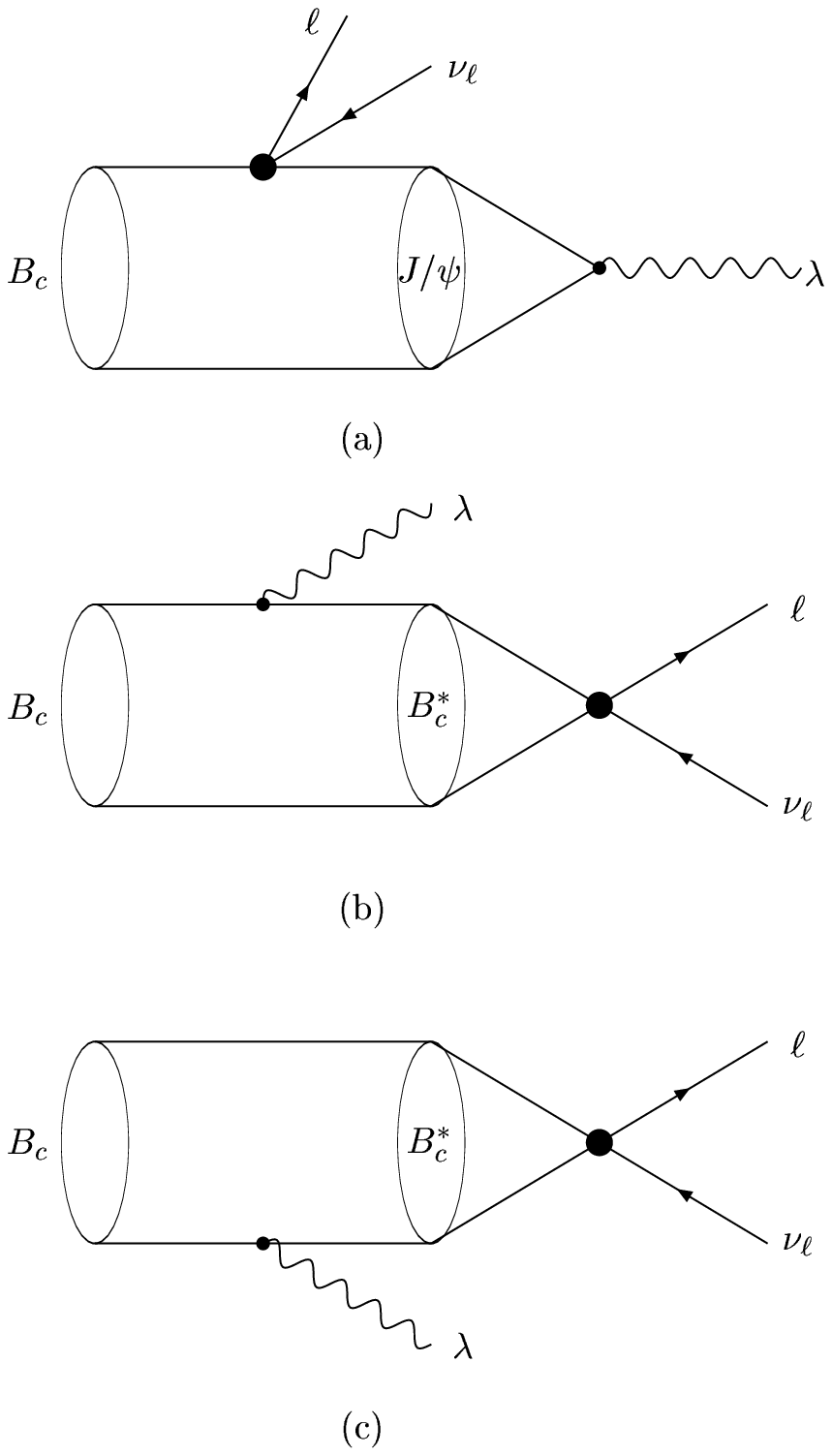, bbllx=178pt,bblly=235pt,bburx=425pt,bbury=673pt,
width=10cm,angle=0}
\caption{\bf Diagrams: (a) $J/\psi$ as an 
intermediate state; (b) and (c) $B_{c}^{*}$ as an intermediate state.}
%\label{fig}
\end{center}
\end{figure}

\begin{figure}\begin{center}
   \epsfig{file=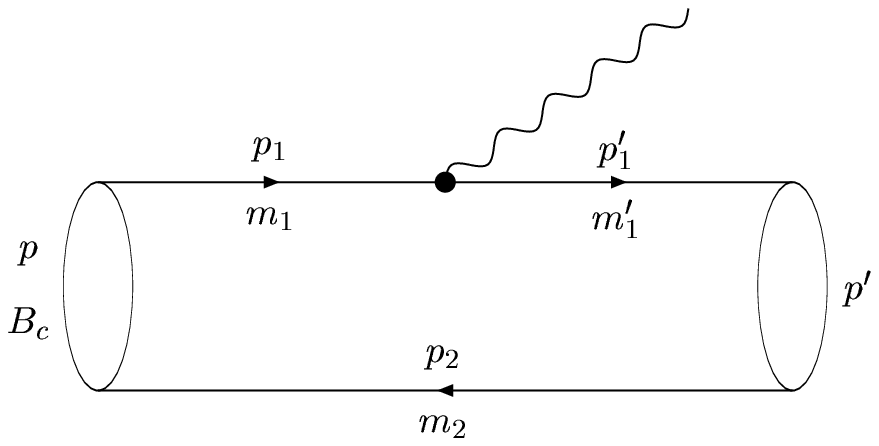, bbllx=178pt,bblly=508pt,bburx=432pt,bbury=640pt,
width=10cm,angle=0}
\caption{\bf A Feynman diagram corresponding to the relevant weak or electromagnetic current  
 sandwiched by the $B_{c}$ meson 
and a suitable single-particle state.}
%\label{fig}
\end{center}
\end{figure}

\begin{figure}\begin{center}
   \epsfig{file=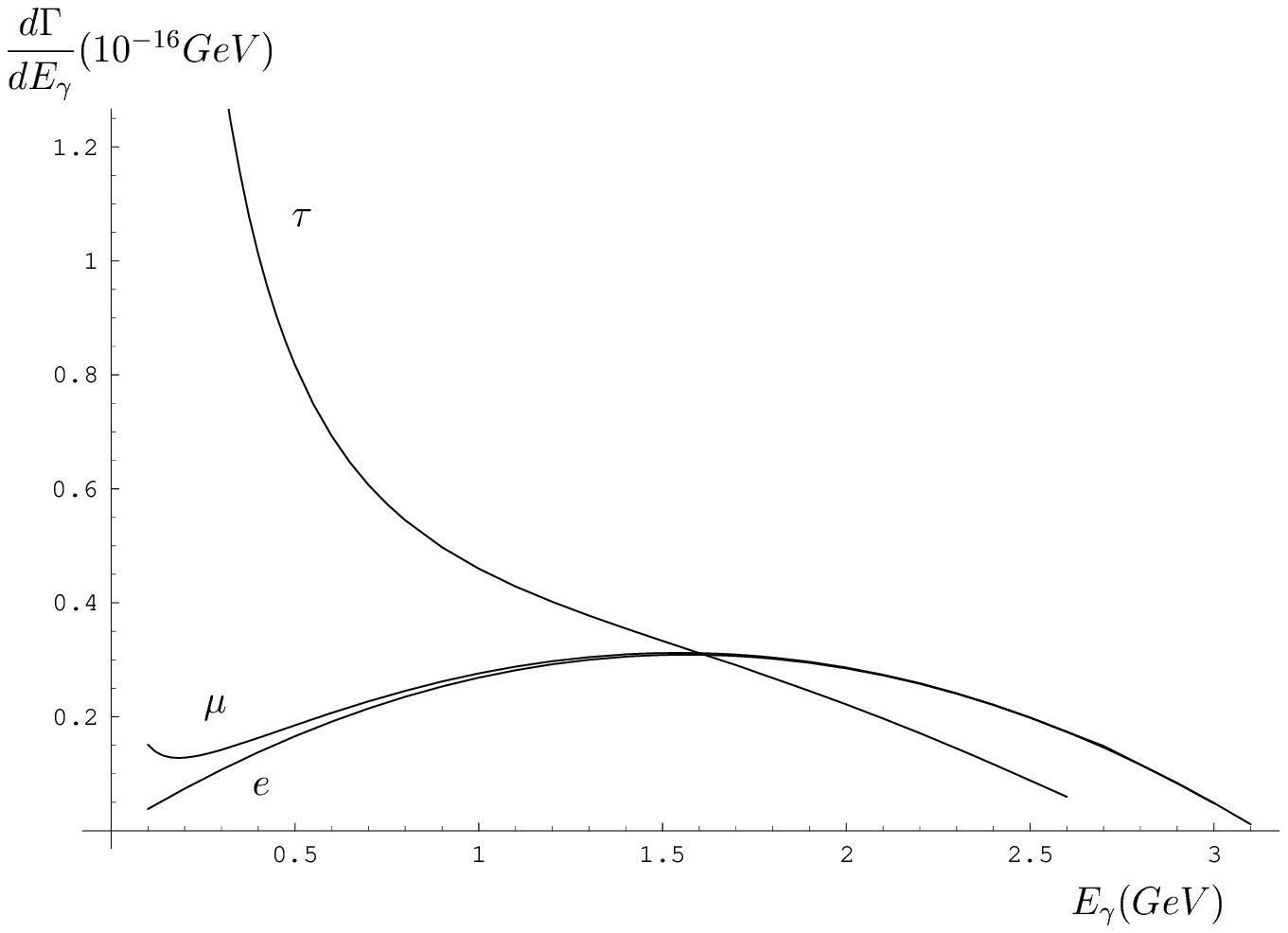, bbllx=160pt,bblly=300pt,bburx=550pt,bbury=580pt,
width=10cm,angle=0}
\caption{\bf Photon energy spectra of radiative decays  $B_{c}\longrightarrow\ell{\nu_{\ell}}
\gamma(\ell=e, \mu, \tau)$.}
%\label{fig}
\end{center}
\end{figure}
\end{document}